# Driving the field-free Josephson diode effect using Kagome Mott insulator barriers

Michiel P. Dubbelman[1,2,*], Heng Wu[1,2,3,*,+], Joost Aretz[4], Yaojia Wang[1,2,3], Chris M. Pasco[5], Yuzhou Zhao[6,7], Trent M. Kyrk[1,2], Jihui Yang[7], Xiaodong Xu[6,7], Tyrel M. McQueen[5], Malte Roesner[4], Mazhar N. Ali[1,2+]

[1]Department of Quantum Nanoscience, Faculty of Applied Sciences, Delft University of Technology, Lorentzweg 1, 2628 CJ, Delft, The Netherlands

[2]Kavli Institute of Nanoscience, Delft University of Technology, Lorentzweg 1, 2628 CJ, Delft, The Netherlands

[3]Quantum Solid State Physics, KU Leuven, Leuven 3001, Belgium

[4]Institute for Molecules and Materials, Radboud University, 6525 AJ Nijmegen, The Netherlands

[5]Department of Chemistry, Department of Materials Science and Engineering, William H. Miller III Department of Physics and Astronomy, Johns Hopkins University, Baltimore, Maryland 21218, USA

[6]Department of Physics, University of Washington, Seattle, Washington 98195, USA

[7]Department of Materials Science and Engineering, University of Washington, Seattle, Washington 98195, USA

[*]These authors contributed equally: Michiel Dubbelman, Heng Wu

[+]e-mail: wuhenggcc@gmail.com, m.n.ali@tudelft.nl

**ABSTRACT**

Josephson junctions (JJs), devices consisting of two superconductors separated by a barrier, are of great technological importance, being a cornerstone of quantum information processing. Classical understanding of superconductor-insulator-superconductor JJs is that conventional insulator's properties, other than magnetism, do not significantly influence the junction's behavior. However, recent work on quantum material (QM) JJs – using Mott insulator $Nb_3Br_8$ – resulted in magnetic field-free non-reciprocal superconductivity, termed the Josephson diode effect (JDE), implying the QM's intrinsic properties can modulate superconductivity in non-trivial ways. To date, the underlying mechanism and dependence of the JDE on correlation strength (U/t) has not been elucidated. Here we fabricate QMJJs using correlated Kagome insulators with varying U/t, $Nb_3X_8$ (X=Cl, Br, I), observing a decreasing trend of the field-free JDE with $Nb_3Cl_8$ reaching ~48% efficiency, $Nb_3Br_8$ ~6%, and $Nb_3I_8$ having no discernible JDE, matching the trend of decreasing U/t from Cl to I and suggesting correlation in insulators drives the field-free JDE.

**INTRODUCTION**

Josephson junctions consisting of two superconductors connected by a weak link are of increasing importance due to their application as qubits in quantum computing, logic gates in superconducting computing, their central role in quantum sensors, and more (1–4). One of the major reasons JJs have become so ubiquitous is that they can be fabricated from a wide range of different types of materials; both the superconductor and the weak link can be changed. The superconductor can be type-I, type-II, unconventional, or high $T_c$, and the weak link can range from an insulator (SIS), to a metal (SNS) or a combination of different materials (e.g. SINIS), to fit the necessary application.

In all of these different architectures, the supercurrent across the junction follows the Josephson relations, which are periodically dependent on the phase difference between the superconductors (5, 6). In SIS junctions in particular, the insulating barrier serves as a simple tunnelling barrier, and its own intrinsic properties are generally neglected. A widely accepted exception is magnetism where, for example, in a JJ with a ferromagnetic insulator, localized magnetic moments produce a strong exchange field which influences the spin state of the tunneling Cooper pairs, causing phase shifts (7, 8) and non-traditional current-phase relations. This can result in complex critical current behavior as a function of external magnetic field (7) including a magnetic field dependent Josephson diode effect (9–13), where the critical current through the JJ is non-reciprocal. These magnetic field dependent JDE always show a switching of the direction of the JDE with at a particular magnetic field where the positive and negative critical currents are equal. This crossing occurs at a finite field for JDE devices with internal magnetism (*14*, *15*), and at zero field for JDE devices without internal magnetism (*10*, *13*, *16*, *17*).

On the contrary, the magnetic *field-free* Josephson Diode effect (18–21) hosts a supercurrent in one direction but a resistive current in the other *without* the presence of an applied magnetic field. This is analogous to the semiconducting diode and has applications in superconducting electronics including rectifiers, gyrators, circulators, and more (22–25). The field-free JDE was first realized in a van der Waals heterostructure of NbSe$_2$/Nb$_3$Br$_8$/NbSe$_2$ (18), where NbSe$_2$ is a 2D superconductor, and Nb$_3$Br$_8$ is an insulating barrier and a member of the Kagome material family Nb$_3$X$_8$ (X=Cl, Br, I) (26). Kagome materials, consisting of corner shared hexagons and triangles, have recently attracted a lot of attention due to their ability to simultaneously host flat bands, a Dirac point, and van Hove singularities (27–31) in their electronic structures. Compared to the ideal Kagome lattice, the Nb$_3$X$_8$ family possesses the breathing Kagome lattice (detailed below), which opens a gap at the Dirac point, allowing for isolated flat bands (32–35) which cause Nb$_3$Br$_8$ to be a non-magnetic, strongly correlated (Mott) insulator (34, 36–42). A possible origin of the observed field-free JDE (with a non-magnetic barrier material) is that the correlation and layer-dependence of Nb$_3$Br$_8$ breaks inversion and time-reversal symmetry allowing for non-reciprocal superconductivity (43, 44). This suggests a new way in which the barrier material can influence the properties of SIS junctions and to date, Nb$_3$Br$_8$ is the only barrier material known to exhibit such a field-free JDE.

In this work, we explore the effect of varying correlation strength of the barrier on the JDE by measuring the transport properties of the full set of Nb$_3$X$_8$ (X=Cl, Br and I) JJs using NbSe$_2$ as the superconductor. Among them, the JJs with Nb$_3$Cl$_8$ and Nb$_3$Br$_8$ show the field-free JDE, while the JJ with Nb$_3$I$_8$ does not. Additionally, the Nb$_3$Cl$_8$ JJ exhibits the highest diode efficiency among them, reaching ~48% at $T = 260$ mK and zero field. Additionally, we observe an anomalous, non-Ambegaokar-Baratoff, $I_C$ dependence with temperature in the Nb$_3$Cl$_8$ JJs. Recent theoretical calculations show a trend of decreasing electron correlation strength from Nb$_3$Cl$_8$ to Nb$_3$Br$_8$ to Nb$_3$I$_8$, (36), with Nb$_3$Cl$_8$ being a strongly correlated Mott insulator and Nb$_3$I$_8$ being a weakly correlated or trivial insulator. This matches our observed trend of the field-free JDE, showcasing that correlation in the insulating barrier can modulate superconductivity across the junction, challenging conventional JJ models and opening new opportunities for both fundamental research as well as technological innovation in superconducting electronics.

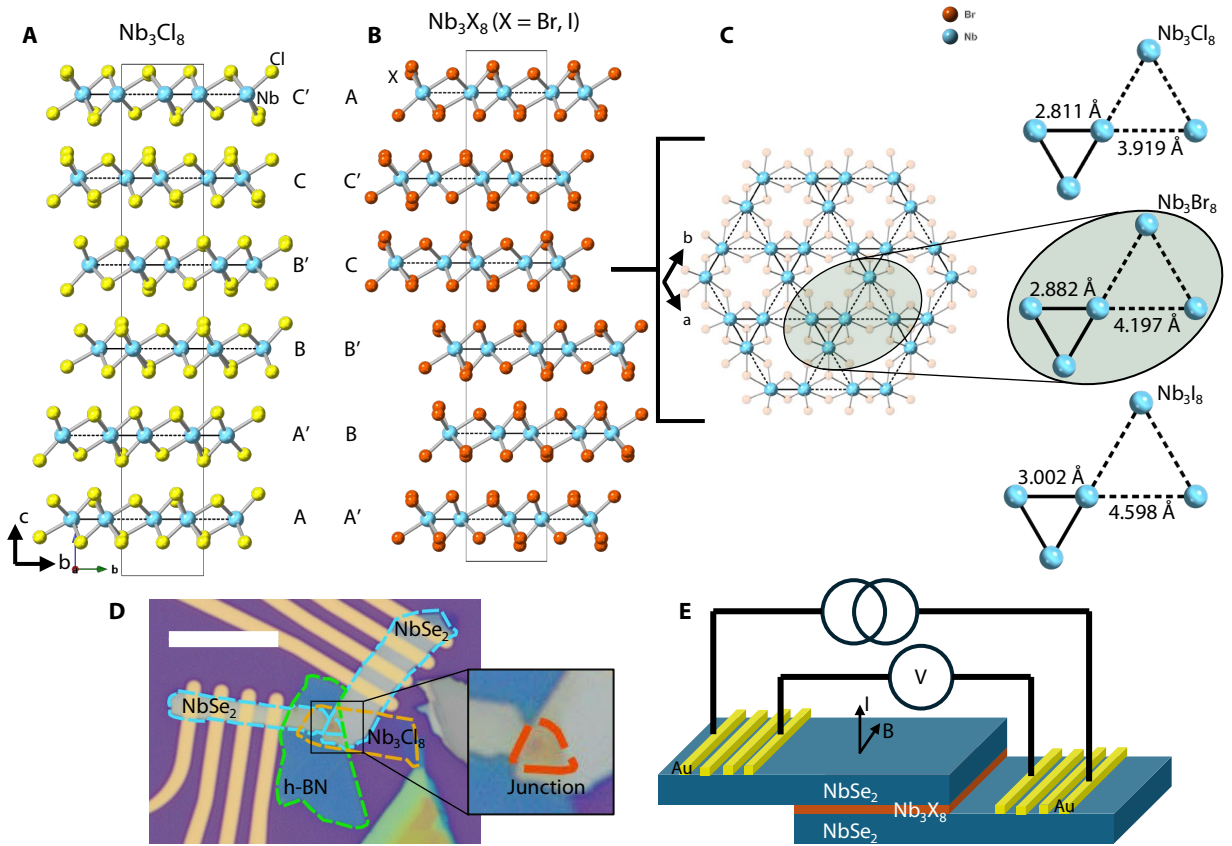

**Fig. 1. Crystal structure of Nb$_3$X$_8$ and a typical device image.** (**A**) Crystal structure of R3 Nb$_3$Cl$_8$. It has six-layer stacking where the prime (e.g. A') and non-prime (e.g. A) layers are inversion symmetric. The size of the niobium trimers change slightly between the prime and non-prime layer from 2.801 Å to 2.821 Å. (**B**) Schematic picture of the crystal structure of Nb$_3$Br$_8$/Nb$_3$I$_8$. It has six-layer stacking where the prime and non-prime layers are inversion symmetric. (**C**) Schematic picture of single layer Nb$_3$X$_8$ with the niobium highlighted. The breathing kagome lattice is clearly visible with the triangles connecting at the corners. The popout shows the niobium unit cell for all three Nb$_3$X$_8$. For Nb$_3$Cl$_8$ the average of the two different niobium distances is shown. The trimers increase clearly from Nb$_3$Cl$_8$ to Nb$_3$I$_8$. (**D**) Microscope picture of a representative device. The NbSe$_2$/Nb$_3$Cl$_8$/NbSe$_2$ junction is capped with h-BN. The scale bar is 10 μm. In the popout a zoom in on the junction is shown. The junction area is 2.3 μm$^2$. (**E**) Schematic picture of the four-probe measurement setup. The current goes vertical through the junction and, when applied, the magnetic field is in plane of the flakes.

## RESULTS

The niobium halides are van der Waals materials where molecular layers stack into a six-layer repeating pattern at low temperature (Fig 1A, 1B). The molecular layers comprise a breathing Kagome lattice, where niobium atoms form a corner sharing triangular lattice (trimer) with two different triangle sizes as shown in Fig. 1C. For $Nb_3Br_8$ and $Nb_3I_8$ the low temperature space group is reported to be $R\bar{3}m$ (see Fig. 1B) (45–49), whereas that of $Nb_3Cl_8$ is still under debate, with proposed space groups including $R3$ (50, 51), $R\bar{3}m$ (26, 46, 52, 53), and $C2/m$ (54). The niobium trimer is smallest in $Nb_3Cl_8$, with bond distances of 2.801 Å and 2.821 Å for layers with an apostrophe mark and without one, respectively. This distance increases to 2.882 Å for $Nb_3Br_8$ and 3.002 Å for $Nb_3I_8$. Similarly, the intralayer and interlayer distances increase from Cl to I.

For the JJs, $NbSe_2$ is used as the van der Waals superconductor due to its relatively high critical temperature and current. An optical image of a typical $NbSe_2/Nb_3Cl_8/NbSe_2$ device with h-BN as the capping layer is shown in Fig. 1D. In order to investigate the transport properties of the JJ with the $Nb_3X_8$ barrier, a four-probe method is adopted with the schematic shown in Fig. 1E, where the outer two electrodes are used as the applied current source, and the inner two electrodes are used as voltage probes. Additionally, since the devices are all vertical JJs (current applied out of plane), applied magnetic fields are in-plane, perpendicular to the overlapped area as illustrated in Fig. 1D.

We first investigated the JDE behavior in a $NbSe_2/Nb_3Cl_8/NbSe_2$ heterostructure, denoted as device #Cl1. Fig. 2A shows the V-I curve measured at T = 200 mK with a positive sweep (applied current from 0-p, p-0, 0-n and n-0) (18) without any external magnetic field. An obvious difference between the positive ($I_{C+}$) and negative critical current ($I_{C-}$) is visible, showing the presence of a large field-free JDE. To explore the temperature dependent JDE behavior, the V-I curves at temperatures ranging from T = 90 mK to T = 5 K were measured and the calculated differential resistance (dV/dI) as a function of applied current and temperature are plotted in a color map shown in Fig. 2C.

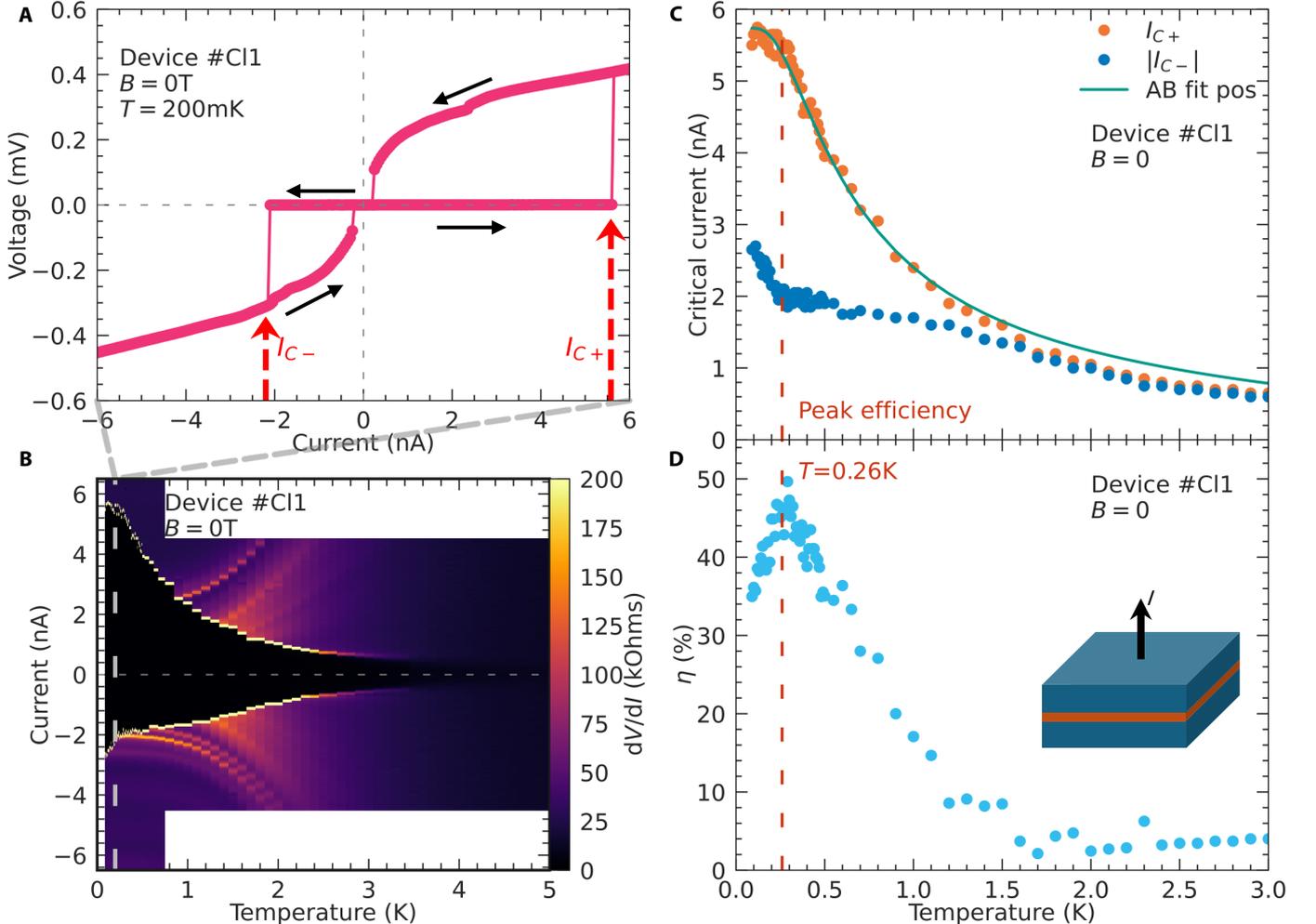

**Fig. 2. V-I characteristics at different temperature at zero field of $NbSe_2/Nb_3Cl_8/NbSe_2$.** (**A**) V-I measurement taken at zero field and T = 200 mK of device #Cl1. The black arrows indicate the current sweep direction in each part of the measurement. The red arrows show where the current breaks the superconductivity, which is vastly different comparing the positive and negative critical current. (**B**) Calculated dV/dI map versus temperature. Each temperature is one V-I measurement. (**C**) The extracted positive and absolute negative critical current ($I_{C+}$, $|I_{C-}|$) versus temperature. The diode effect appears roughly at T = 1.6 K. The orange dashed line indicates the peak efficiency at T = 260 mK. The teal line is the fit to the AB-model. (**D**) The superconducting diode efficiency $\eta$ versus temperature, with the orange line indicating the peak efficiency. The inset shows the current direction in the vertical junction.

Besides the superconducting transitions being the first bright spot on each V-I curve, there are more transitions appearing on the V-I curve above the superconducting transition in the normal state of the JJs. These additional transitions (visible as kinks in the p-0 and n-0 parts of the V-I curves in Fig. 2A, S1-3) are not symmetric with I; more transitions are visible at low temperature in $I_-$ than in $I_+$ and they are not related to the superconducting transition of the NbSe$_2$ flakes, as they are well below the critical current for even thin NbSe$_2$. These may correspond to quasiparticle tunneling or phase slips in these tunneling JJs (6, 55–58), however explaining this normal state behavior is beyond the scope of this work. Both the positive and negative critical currents were extracted from the colormap and their magnitudes plotted in Fig. 2C for better comparison. As can be clearly seen, $I_{c+}$ is larger than $I_{c-}$ when the temperature is below T = 1.6 K, indicating the JDE emerges only when the temperature is relatively lower than $T_c$ of the JJ. The temperature dependence of $I_c$ is fit to the Ambegaokar-Baratoff (AB) model (5, 6, 59):

$$I_c = \frac{\pi}{2eR_n} \Delta(T) \tanh\left(\frac{\Delta(T)}{2k_B T}\right) \quad (1)$$

where e is the elementary charge, $R_n$ is the normal state resistance, $\Delta$ is the superconducting gap, $k_B$ is the Boltzmann constant and T is the temperature. By fitting $R_n$ and $\Delta(0)$ to the positive critical currents, we extract a normal state resistance $R_{n,fit}$=21 kOhm, similar to the measured normal state resistance $R_{n,meas}$=26 kOhm. However the extracted superconducting gap for device #Cl1 (#Cl2) $\Delta_{0,fit}$=75 µeV (63 µeV) is anomalously small; only ~6% (5%) of the reported gap $\Delta_0$=1.26 meV of NbSe$_2$ (60). A comparison between the experimental and freely fit parameters is visible in fig S5. This aligns with predictions of a suppressed critical current in correlated insulator barriers, where the strong electron correlation drives the critical current below AB model predictions (61–64). Additionally, in Fig. 2C, a kink exists at T = 260 mK at $I_{c-}$ which is absent in $I_{c+}$, with an upturn in $I_{c-}$ with decreasing T, completely opposite to the plateauing behavior of $I_{C+}$. These features imply that there may be different tunnelling mechanisms for Cooper pairs for the positive and negative applied current.

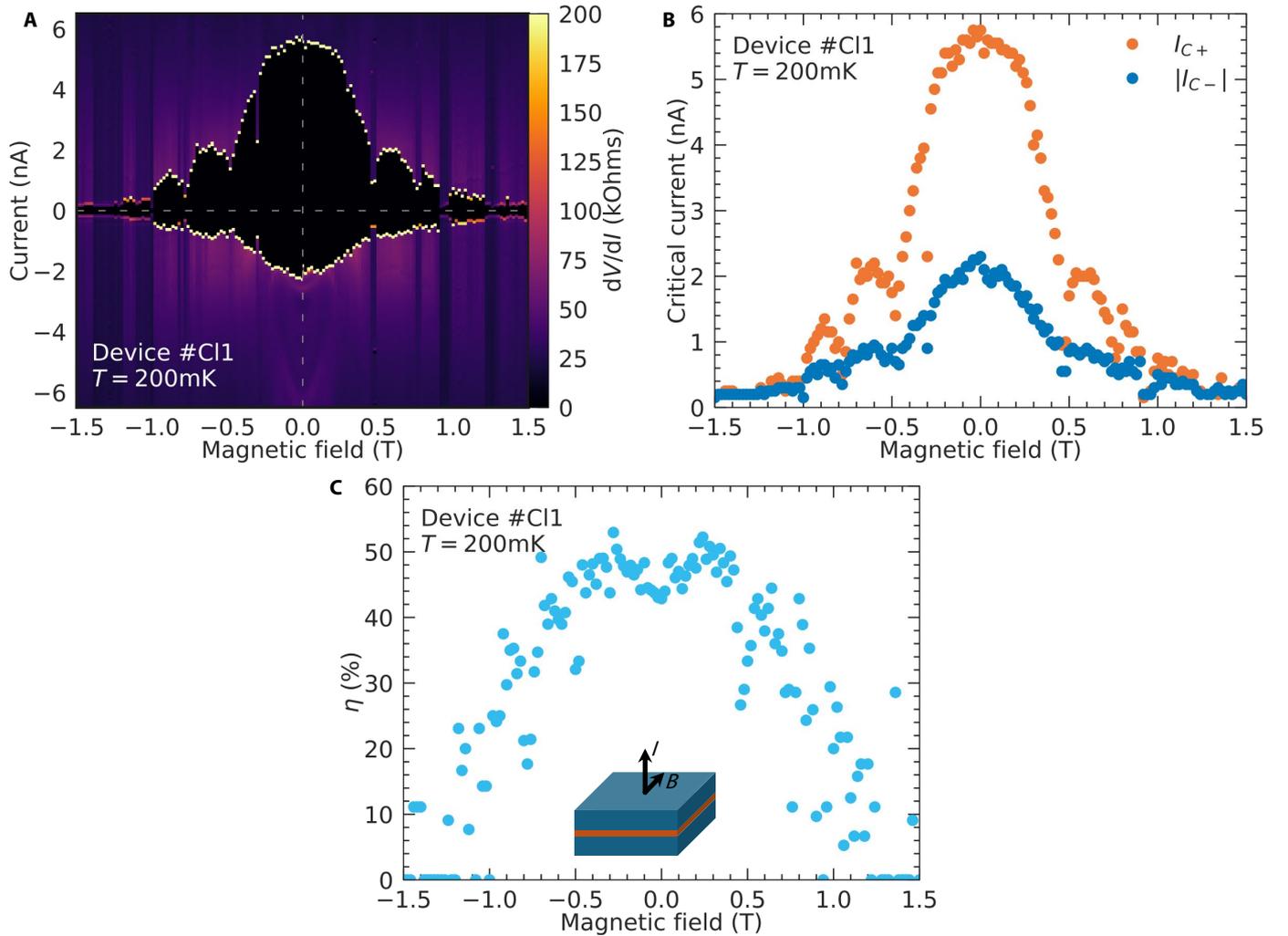

**Fig. 3. Magnetic field dependence of NbSe$_2$/Nb$_3$Cl$_8$/NbSe$_2$ at T = 200 mK.** (**A**) Calculated dV/dI map versus magnetic field. Each field is one V-I measurement. (**B**) The extracted positive and absolute negative critical current ($I_{c+}$, $|I_{c-}|$) versus field. The diode effect is clearly even with respect to magnetic field. $I_{c+}$ is bigger than $|I_{c-}|$ up until +/- 1 T. (**C**) The superconducting diode efficiency η versus field. The maximum efficiency reached is roughly 50 %. The inset shows the current direction in the vertical junction and the magnetic field applied in plane.

The diode efficiency $\eta = (I_{C+} - |I_{C-}|) \div (I_{C+} + |I_{C-}|)$ was calculated as a function of temperature and plotted in Fig. 2D. The kink in $I_{C-}$ but absent in $I_{C+}$ results in a peak in $\eta$ at $T = 260$ mK, which reaches up to 48% in absence of magnetic field. Similarly, non-AB temperature dependent behavior was seen in device #Cl2 (Fig. S4).

To explore the magnetic field dependent JDE behavior, V-I curves were taken at different external magnetic fields at a constant temperature of 200 mK; several typical curves are plotted in supplementary materials fig. S2A-B. Similar to previous measurements, the differential resistances were calculated and plotted in a colormap as functions of applied magnetic field and current, shown in Fig. 3A. The magnetic field dependence of $I_{C+}$ follows a Fraunhofer-like pattern, indicating Josephson coupling in the NbSe$_2$/Nb$_3$Cl$_8$/NbSe$_2$ heterostructure. Deviation from the ideal Fraunhofer pattern is expected, originating from the irregular shape of the tunneling area (65, 66). On the contrary, due to the suppressed magnitude, the oscillations in the $I_{C-}$ can barely be observed, again indicating possible direction dependent tunnelling mechanisms of Cooper pairs. Additionally, the normal state resistances vary with different external magnetic fields, as can be seen from the different vertical strips with different colors.

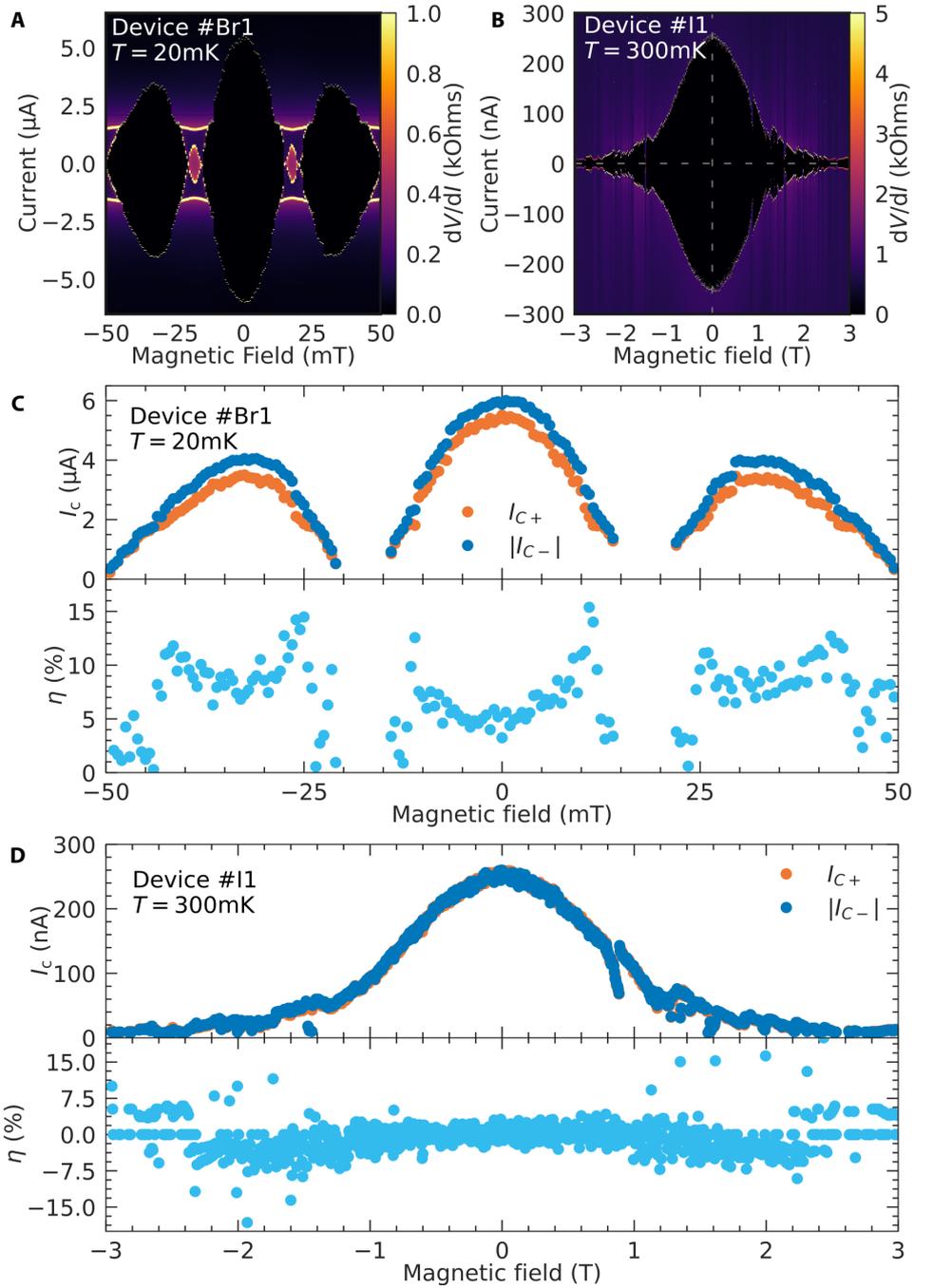

**Fig. 4. Magnetic field dependence of NbSe$_2$/Nb$_3$Br$_8$/NbSe$_2$ & NbSe$_2$/Nb$_3$I$_8$/NbSe$_2$.** (**A**) Calculated dV/dI map versus magnetic field for NbSe$_2$/Nb$_3$Br$_8$/NbSe$_2$ device #Br1 at $T = 20$ mK. Each field is one V-I measurement. (**B**) Calculated dV/dI map versus magnetic field for NbSe$_2$/Nb$_3$I$_8$/NbSe$_2$ device #I1 at $T = 300$ mK. Each field is one V-I measurement. (**C**) Upper panel: The extracted positive and absolute negative critical current ($I_{C+}, |I_{C-}|$) versus field for device #Br1. The diode effect is clearly even with respect to magnetic field. The diode effect turns of with the disappearing superconductivity and reappears when the superconducting state reappears. Lower panel: The superconducting diode efficiency $\eta$ versus field. The maximum efficiency reached is roughly 15% (**D**) Upper panel: The extracted positive and absolute negative critical current ($I_{C+}, |I_{C-}|$) versus field for device #I1. The critical currents almost perfectly line up. Lower panel: The superconducting diode efficiency $\eta$ versus field. It hovers around 0% around 0 T with only the noise contributing.

Several typical V-I curves are plotted in the supplementary materials Fig. S3A-B, where different tendencies of voltage behaviors after breaking the superconductivity can be clearly observed. $I_{C+}$ and $I_{C-}$ were extracted and plotted in Fig. 3B, where $I_{C+}$ is larger than $|I_{C-}|$ from -1 T to 1 T, revealing the JDE persists over a large range of magnetic field. The symmetric $I_C$ vs $B$ curves also demonstrate the existence of the JDE at zero magnetic field. As shown in Fig. 3C, $\eta$ reaches ~45% at zero field and 200mK, and increases to 50% at $B = 300$ mT, after which it decreases to 0 by $B \approx 1.2$T.

To investigate the JDE evolution with varying barrier correlation strength, Josephson junctions with $Nb_3Br_8$ and $Nb_3I_8$ were fabricated and measured as well. Fig. 4A shows the differential resistance colormap of a JJ with $Nb_3Br_8$ (device #Br1). The oscillations in the Fraunhofer-like colormap indicates Josephson coupling in the device. Though the differential resistances between the lobes of the colormap exhibit strange behaviors (small pockets of resistive state appear in between the superconducting lobes), we can still observe the JDE in the main lobes. We extracted the $I_{c+}$ and $I_{c-}$ and calculated $\eta$ for three lobes, as shown in Fig. 4C, where the JDE can be clearly observed. In each lobe, the behavior of the $\eta$ resembles the behavior in device #Cl1; that is, each lobe shows the JDE in the center with a quick drop to 0 by the edge of the lobes. The symmetric $\eta$ vs. magnetic field (like with $Nb_3Cl_8$) again demonstrates the presence of the field-free JDE, but the $\eta$ with $Nb_3Br_8$ is only ~6 % at zero field, and reaches up to ~15 % at 11 mT applied field, in solid agreement with the previously reported $NbSe_2/Nb_3Br_8/NbSe_2$ JJ behavior (*18*).

The differential resistance of $NbSe_2/Nb_3I_8/NbSe_2$ device (device #I1) was also measured and plotted in Fig 4B. However, even though a Fraunhofer-like pattern is again observed, there is no obvious difference between $I_{c+}$ and $I_{c-}$, meaning no field-free or field dependent JDE is seen. This can be further confirmed by the overlapped $I_{c+}$ and $I_{c-}$ and the near-zero $\eta$ shown in Fig. 4D.

## **DISCUSSION**

Our results show a large field-free JDE in $NbSe_2/Nb_3Cl_8/NbSe_2$ of $\eta$ = 48 % that decreases across the halogen series with $\eta$ = 6 % for $NbSe_2/Nb_3Br_8/NbSe_2$ and no JDE in $NbSe_2/Nb_3I_8/NbSe_2$ JJs. Given that these devices are constructed with similar thicknesses of $NbSe_2$, halide barrier, and device geometries (see supplement), the major difference between them is the nature of the barrier material, implying these JDE behaviors are driven by the intrinsic properties of the $Nb_3X_8$. Interestingly, the evolution of electron correlation in the $Nb_3X_8$ family matches the evolution of the field-free JDE; that is, with a Mott insulating barrier, the field-free JDE is observed, and with a non-Mott insulating barrier, it is absent.

Firstly, ARPES measurements in combination with theoretical calculations (*32–34, 36, 39, 48*), confirm $Nb_3Cl_8$ and $Nb_3Br_8$ bulk crystals to be Mott insulators, which according to recent STM experiments should persist down to the few-layer regime in $Nb_3Cl_8$ (*35*). While $Nb_3I_8$ has a qualitatively similar band structure to $Nb_3Cl_8$ and $Nb_3Br_8$ (*32*), recent ab-initio calculations predict that the Mott insulating behavior is absent until the monolayer regime (*36*). Furthermore, these calculations show that the onsite coulomb repulsion, $U$, vs the hopping of the electrons, $t$, decreases from $Nb_3Cl_8$ to $Nb_3Br_8$ to $Nb_3I_8$. Previous theoretical work on SIS junctions with Mott insulating barriers and broken inversion symmetry also suggested a field-free JDE (*43, 44*). Taken together, these phenomena suggest that the correlation strengths of the barriers drive the field-free JDE.

When comparing the field-free JDE reported here to other reported field-free platforms, several differences stand out; specifically when compared to the $NbSe_2/Nb_3Cl_8/NbSe_2$ junctions. We observe a JDE over a wide range of magnetic field, being able to apply up to 1 T, compared to the highest other field reported at 200mT (*20*). Furthermore, the diode efficiency, reaching 48% at zero field, is twice the value of the highest reported field-free superconducting diode platform, at 22% reported in BSCCO twisted films (*67*). Additionally, the symmetric behavior of the $Nb_3Cl_8$ and $Nb_3Br_8$ JDE with an applied magnetic field is so-far unique amongst other JDE platforms. In non-magnetic, non-correlated insulator junctions showing a field dependent JDE, a crossing point between $I_{c-}$ and $I_{c+}$ is always visible at zero applied field (*10, 13, 16, 17*); this aligns with the idea that without a magnetic field these systems do not break both inversion and time reversal symmetry and cannot host a non-reciprocal response like a JDE. In magnetic, non-correlated insulator platforms that use an intrinsic ferromagnetic state to create a field-free JDE, the magnetic field dependence shows a crossing point that is shifted from zero-field by the magnitude of the internal field from the ferromagnetic state (e.g. $Fe_3GeTe_2$ JJs (*14*)). To the best of the author's knowledge, only $Nb_3Cl_8$ and $Nb_3Br_8$ junctions show a magnetic field dependence with no crossing point, aligning with the claim of these being non-magnetic or antiferromagnetic, so no net magnetization is present in the material.

This leads into the question of how time reversal symmetry breaking is achieved in the $NbSe_2/Nb_3X_8/NbSe_2$ JJs. Inversion symmetry breaking is naturally achieved due to the asymmetric bottom and top interfaces between $NbSe_2$ and $Nb_3X_8$, but the requirement and mechanism of time-reversal symmetry breaking is still under debate (*18, 44*). To date, no intrinsic magnetic response has been observed in $Nb_3X_8$ at low temperatures (*26, 45, 46, 50, 54*), but their monolayers are predicted to be either ferromagnetic with out of plain momentum (*68, 69*) or antiferromagnetic for bilayers with two layers forming a singlet state (*33, 36*). Further experimental studies, employing highly sensitive local magnetic techniques, such as muon spectroscopy, are necessary to elucidate the low-temperature magnetic states of $Nb_3X_8$. On the other hand, Nagaosa and others have proposed that out of equilibrium, correlation and dissipation processes can also break time reversal symmetry (*70–74*), causing non-reciprocity in superconducting devices, and alleviating the need to have magnetism involved. To elucidate the microscopic origin, further theoretical and experimental efforts are needed to unequivocally demonstrate the exact relationship between the correlation, magnetism, and the JDE.

In summary, our work demonstrates the presence of the field-free Josephson diode effect in $NbSe_2/Nb_3Cl_8/NbSe_2$ and $NbSe_2/Nb_3Br_8/NbSe_2$, and its absence in $NbSe_2/Nb_3I_8/NbSe_2$ JJs, matching the trend of decreasing correlation strength with heavier halide barrier. This observation showcases that the intrinsic properties of the breathing kagome lattice in $Nb_3X_8$ can be used to produce a robust, but as yet not fully understood, field-free JDE, challenging conventional models of SIS Josephson junctions and opening new opportunities for both fundamental research into the interplay between superconductivity and electron correlation, as well as technological application in superconducting electronics. For example, future QMJJs with correlated barriers and in-situ correlation strength tuning (perhaps via gate control) could provide an avenue for creating field-free tunable Josephson diodes or transistors.

During the preparation of this manuscript, the authors became aware of related work by Wu et al. (*75*) where they show a similar field-free Josephson diode effect in $Nb_3Cl_8$ junctions and study the effect of barrier thickness.

## MATERIALS AND METHODS

The $Nb_3Cl_8$ and $Nb_3Br_8$ crystals were grown using chemical vapor transport as in (*26*) and for the $Nb_3I_8$ as in (*32*). $NbSe_2$ was bought from HQ Graphene. We fabricated the $NbSe_2/Nb_3X_8/NbSe_2$ Josephson junctions in a nitrogen filled glovebox to avoid degradation and oxidation. The bottom $NbSe_2$ flake was exfoliated directly onto a $SiO_2$/Si wafer. $Nb_3X_8$ and the top $NbSe_2$ were exfoliated and transferred on top using the polydimethylsiloxane dry transfer method (*76*). The junction was capped by an h-BN flake to prevent degradation when taking it out in atmosphere. Finally, Ti(5 nm)/Au(50-100 nm) electrodes were fabricated using electron beam lithography and evaporation deposition onto the top and bottom $NbSe_2$.

The transport measurements were performed in two systems. The $Nb_3Cl_8$ and $Nb_3I_8$ devices were measured in a Kiutra L-type rapid cryostat with a bipolar magnet. The a.c. measurements were performed using a Zurich instruments MFLI lock-in amplifier with a frequency of 17.77776 Hz. The IVVI rack from TU Delft was used to inject d.c. current and the d.c. voltage was measured using a Keithley 2182a Nanovoltmeter. Magnetic fields were applied in the ab crystal plane, perpendicular to the current in the junctions. A Keithley 6221 AC/DC Current Source Meter was used to inject both d.c. current, and the d.c. voltage was measured using a Keithley 2182a Nanovoltmeter. The transport properties of the $Nb_3Br_8$ device were measured in a BlueFors dilution refrigerator with a base temperature of 20 mK.

**Acknowledgments:** We acknowledge A. Neil and H. El Mrabet Haje for commenting on an early version of this manuscript.
**Funding:**

NWO Talent Programme VENI financed by the Dutch Research Council (NWO) VI. Veni.222.380 (HW)

NWO Talent Programme VENI financed by the NWO, project no. VI.Veni.212.146 (YW)

KU Leuven Special Research Fund STG/24/071 and no.3E250622 (YW)

NWO Talent Programme VIDI financed by the NWO VI.Vidi.223.089 (MNA)

Kavli Institute Innovation Award 2023 (MNA)

Kavli Institute of Nanoscience Delft (HW, MPD, MNA, TMK)

"Materials for the Quantum Age" (QuMat, registration number 024.005.006) which is part of the Gravitation program financed by the Dutch Ministry of Education, Culture and Science (OCW) (MNA, MD, JA, MR)

David and Lucile Packard Foundation and the Johns Hopkins University Catalyst Award (CMP, TMM)

U.S. Department of Energy (DOE), Office of Science, Basic Energy Sciences (BES), under the award DE-SC0018171 (XD, YZ, JY)

**Author contributions:** H.W. and M.N.A. conceived and designed the study. C.M.P. and T.M.M. grew the $Nb_3Cl_8$ and $Nb_3Br_8$ single crystal. Y.Z. X.X. and J.Y. grew the $Nb_3I_8$ single crystal. M.D., H.W. and Y.W. fabricated the devices and performed the measurements. M.D., H.W. and T.M.K. carried out the data analysis. J.A. and M.R. did the theoretical calculation. J.Y., X.X., T.M.M., M.R. and M.N.A are the Principal Investigators. All authors contributed to the preparation of manuscript.

**Competing interests:** The authors declare that they have no competing interests.

**Data and materials availability:** All data needed to evaluate the conclusions in the paper are present in the paper and/or the Supplementary Materials. Additional data related to this paper may be requested from the corresponding authors.

**Correspondence and requests for materials** should be addressed to H.W. and M.N.A.

## Supplementary

The NbSe$_2$ flake thicknesses were in the range from 20-50 nm. To prevent degradation, the junctions were capped with hexagonal boron nitride (h-BN), as shown in a representative device in Figure 1D. The area of the junctions measured ranges from 0.81 to 8 µm$^2$.

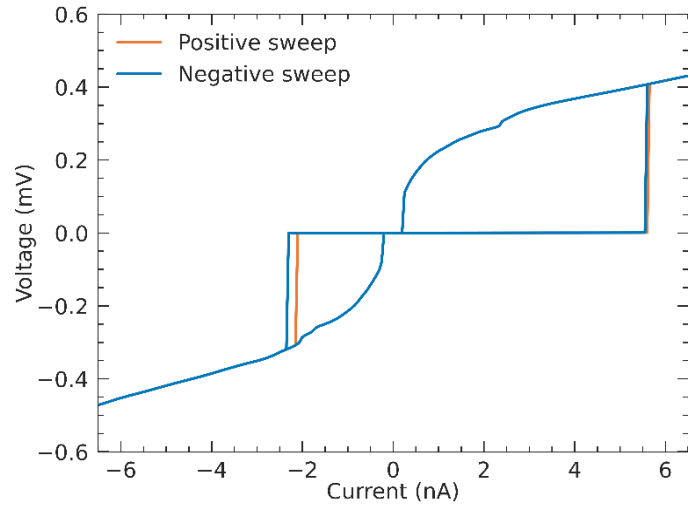

**Fig. S1. Positive and negative *V-I* sweep of device #Cl1.** The positive sweep goes from 0 nA to 6.5 nA, down to -6.5 nA and back to 0 nA. The negative sweep goes the other direction, first going to -6.5 nA. The current sweeps line up well, indicating joule heating effects are of no concern.

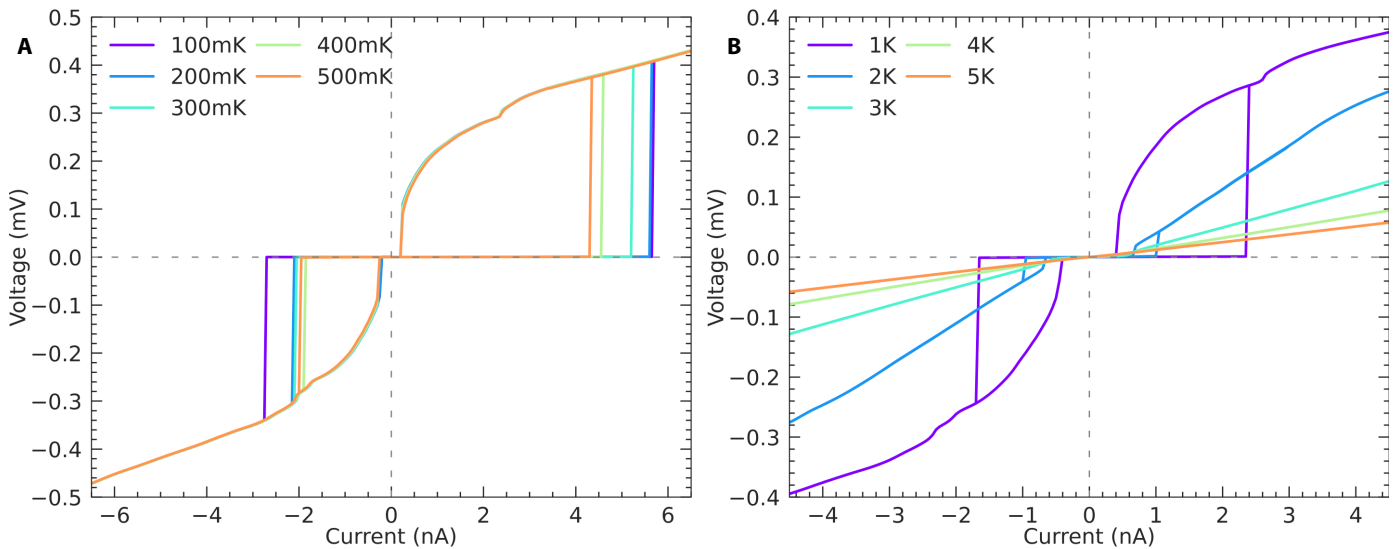

**Fig. S2. Typical *V-I* sweeps at different temperatures of device #Cl1.** (**A**) *V-I* measurements with temperatures from $T = 100$ mK to $T = 500$ mK. The critical current becomes smaller with increasing temperature. The normal state behavior stays the same at all these temperatures. (**B**) *V-I* measurements with temperatures from $T = 1.0$ K to $T = 5.0$ K. The junction becomes fully resistive at higher temperatures.

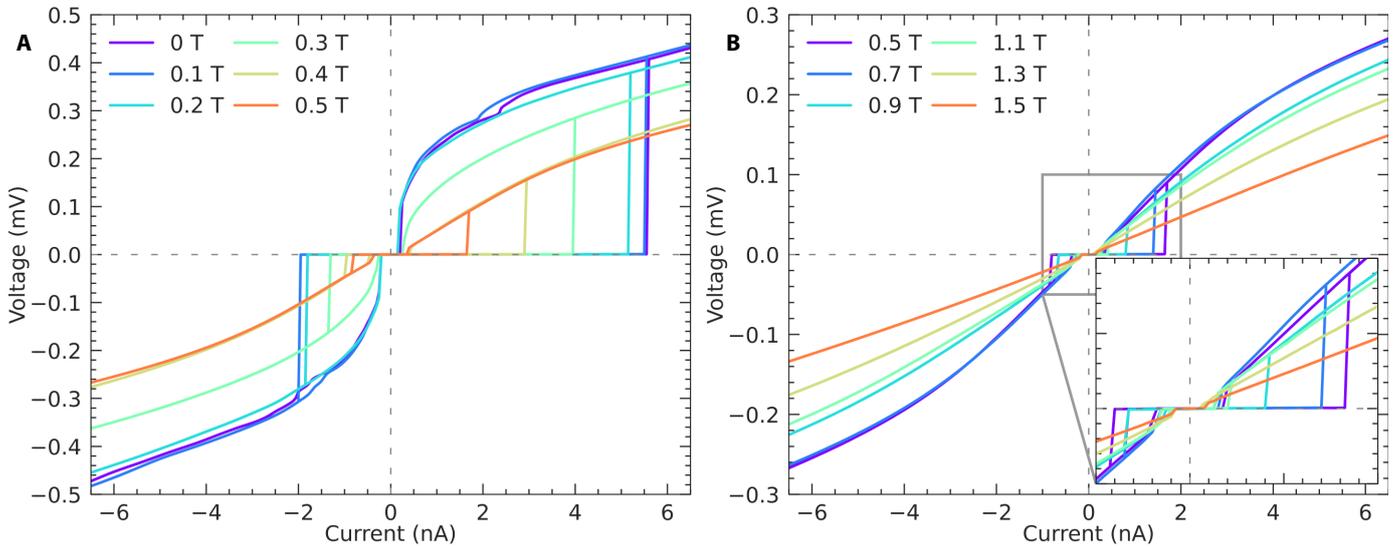

**Fig. S3. Several *V-I* sweeps at different magnetic fields of device #CI1 at 200 mK.** (**A**) *V-I* measurements with fields from 0 T to 0.5 T. The critical current becomes smaller with increasing field. The normal state behavior shows a negative magneto resistive tendency. (**B**) *V-I* measurements with fields from 0.5 T to 1.5 T. Even at 1.5 T the junction has a small superconducting regime at very low currents, as is visible in the zoomed in inset.

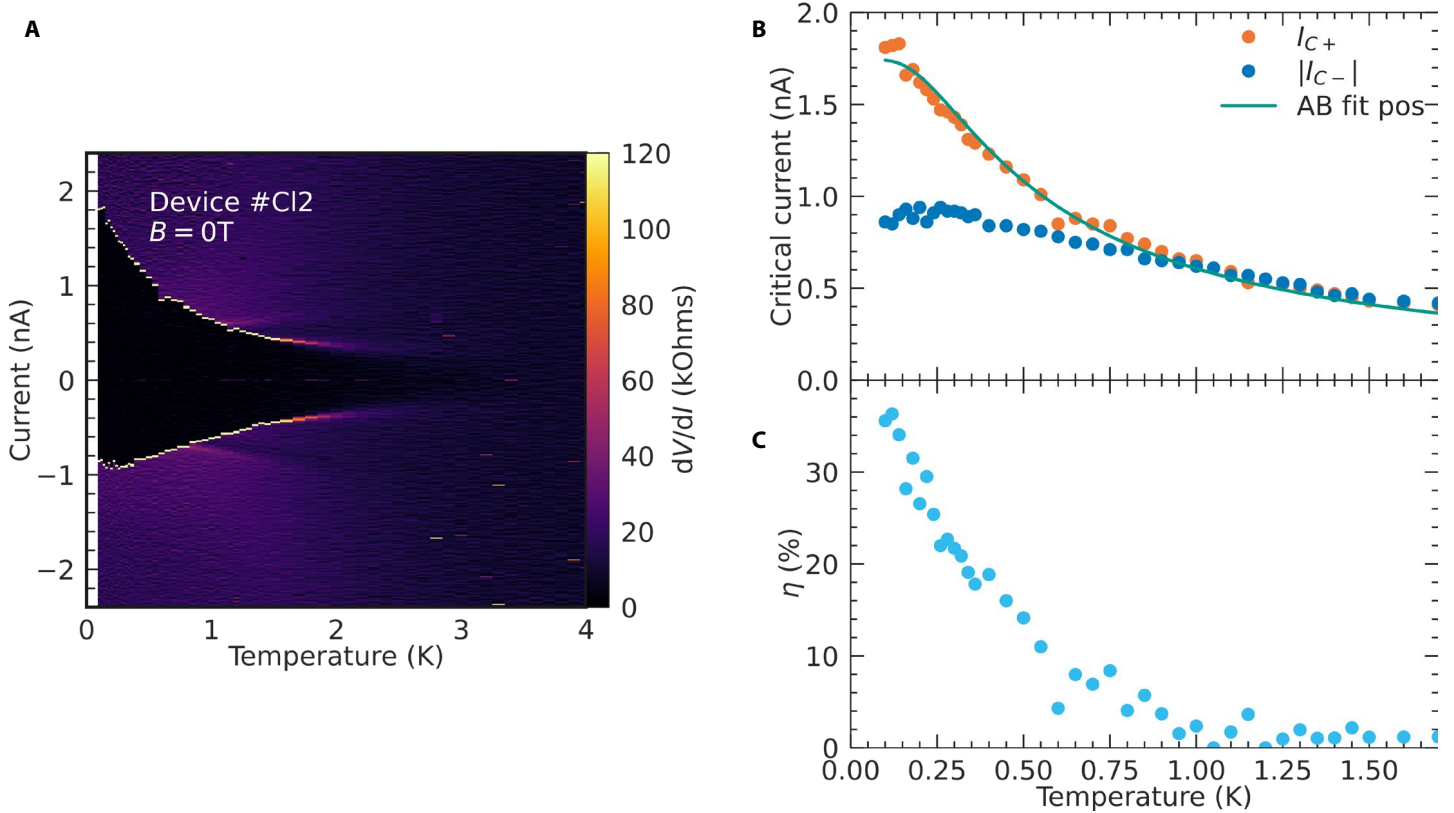

**Fig. S4. V-I characteristics at different temperature at zero field of device #CI2.** (**A**) The calculated d*V*/d*I* colormap versus temperature at zero field is shown. Each temperature is one *V-I* measurement. (**B**) The extracted positive and absolute negative critical current ($I_{c+}$, $|I_{c-}|$) versus temperature. The teal line is the fit to the AB-model. (**C**) The superconducting diode efficiency $\eta$ versus temperature, with now only a rise, since no peak is visible

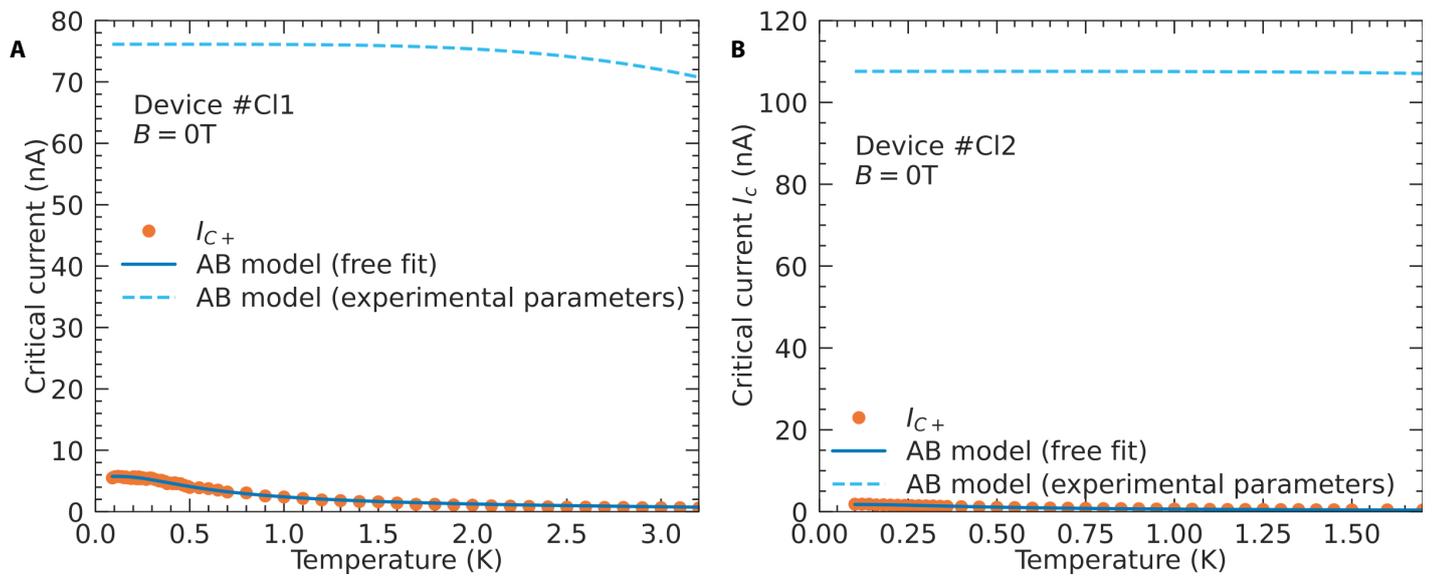

**Fig. S5. Comparing free fit to experimentally determined values of $I_C$ vs $T$.** The figure shows the free fit AB model (dark blue line) versus experimental parameters AB model (light blue dashed line) of device #Cl1 (**A**) and device #Cl2 (**B**), with the data in orange. Clearly, the expected dependence of the AB model from the experimentally determined parameters does not fit the data, while the freely fit parameters fit fairly well. The standard error of regression for the freely fit model for device #Cl1 (#Cl2) is 0.147 nA (0.045 nA).